\documentclass[]{spie}  

\usepackage{amsmath,amsfonts,amssymb}
\usepackage{graphicx}
\usepackage{wrapfig}
\usepackage{tabularx}
\usepackage{capt-of}
\usepackage[colorlinks=true, allcolors=blue]{hyperref}

\title{Characterization of aerogel scattering filters for astronomical telescopes}

\author[a]{Alyssa Barlis}
\author[b]{Stefan Arseneau}
\author[b]{Charles L. Bennett}
\author[a]{Thomas Essinger-Hileman}
\author[c]{Haiquan Guo}
\author[a,d]{Kyle R. Helson}
\author[b]{Tobias Marriage}
\author[a]{Manuel A. Quijada}
\author[e]{Ariel E. Tokarz}
\author[e]{Stephanie L. Vivod}
\author[a]{Edward J. Wollack}
\affil[a]{NASA Goddard Space Flight Center, Greenbelt, MD, USA}
\affil[b]{Johns Hopkins University Department of Physics \& Astronomy, Baltimore, MD, USA}
\affil[c]{Universities Space Research Association, Columbia, MD, USA}
\affil[d]{The University of Maryland, Baltimore County, Baltimore, MD, USA}
\affil[e]{NASA Glenn Research Center, Cleveland, OH, USA}

\authorinfo{Further author information: (Send correspondence to A.B)\\A.B.: E-mail: alyssa.barlis@nasa.gov, Telephone: 1 301 286 6558\\ }

\begin{document} 
\maketitle

\begin{abstract}
We have developed a suite of novel infrared-blocking filters made by embedding scattering particles in a polymer aerogel substrate. Our developments allow us to tune the spectral performance of the filters based on both the composition of the base aerogel material and the properties of the scattering particles. Our filters are targeted for use in a variety of applications, from ground-based CMB experiments to planetary science probes. We summarize the formulations we have fabricated and tested to date, including several polyimide base aerogel formulations incorporating a range of size distributions of diamond scattering particles. We also describe the spectral characterization techniques used to measure the filters’ optical properties, including the development of a mm-wave Fourier transform spectrometer testbed.
\end{abstract}

\keywords{aerogel, filter, blocking, scattering, Fourier Transform Spectroscopy}

\section{INTRODUCTION}
\label{sec:intro}  

Astronomical telescope receivers in the far-infrared and sub-mm spectral ranges require high-quality optical filtration in order to reject out-of-band infrared (IR) thermal loading on highly-sensitive cryogenic detectors. Equally importantly, optical filters must also allow signal in the science band of interest to reach the detectors unimpeded. As receiver instruments under development require increased aperture size ($\sim$1 m in diameter), improvements in detector sensitivity require increasingly stringent optical filtering.

IR-blocking filters can use several physical mechanisms to reject out-of-band radiation, including reflection, absorption, and scattering. Reflective metal-mesh filters are widely used and offer effective IR rejection.\cite{2006SPIE.6275E..0UA} When incorporated into a stack of multiple filters, however, metal-mesh filters provide diminishing returns as it becomes difficult to direct radiation that is reflected multiple times away from the receiver entirely.\cite{2012SPIE.8452E..3IS} Absorptive filters including those made from alumina, nylon, and polytetrafluoroethylene can conduct radiation away from cold focal plane stages, but this mechanism proves challenging with increasing filter diameter because the filters themselves tend to heat up and re-radiate in all directions. Absorptive filters also typically require anti-reflection (AR) coating, which adds complexity to filter fabrication, limits filter bandwidth, and can pose mechanical
challenges. Often a combination of filter types is used to meet the unique requirements of a specific instrument. \cite{1986ApOpt..25..565H,2017ApOpt..56.5349M,1995ApOpt..34.7254B,1974ApOpt..13..425A}

\begin{figure}[h]
\centering
  \includegraphics[width=0.7\textwidth]{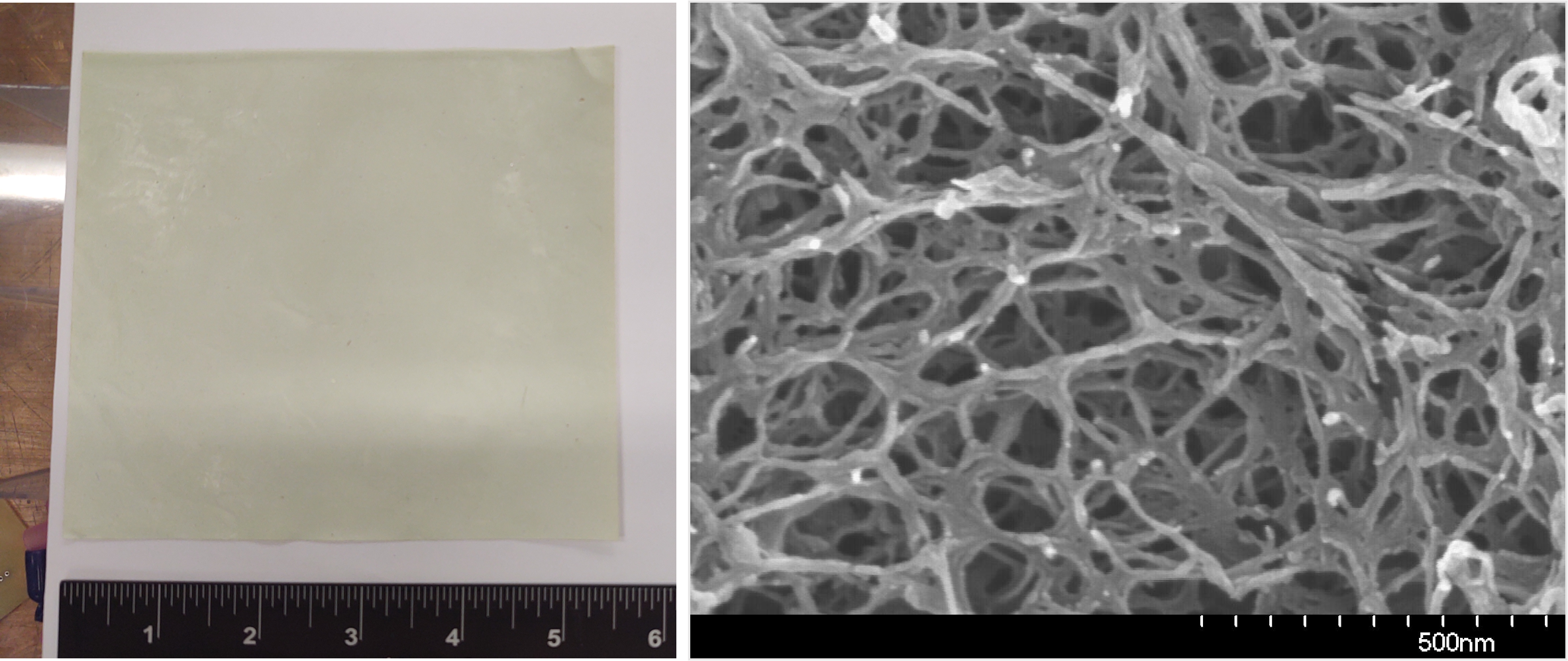}
  \caption{\label{fig:images} Aerogel scattering filter test sample fabricated as a film sheet. This sample consisted of an aerogel backbone of BAPB with DMBZ/BPDA/TAB embedded with 40 mg/cc 40--60 $\mu$m particles, 20 mg/cc 10--20 $\mu$m particles, and 20 mg/cc 3--6 $\mu$m particles. The film density is 0.2 g/cm$^3$ with 90.6\% porosity, and has a thermal conductivity of 25.1 mW m$^{-1}$ K$^{-1}$. Left: The full sample sheet, which takes on a yellow hue due to the polyimide aerogel base. Right: SEM image of the aerogel pore structure, characterized by very large internal surface area.}
\end{figure}

As an alternative to the above approaches, we have developed a suite of aerogel blocking filters by embedding diamond scattering particles in a base polyimide aerogel matrix. Their optical performance can be tuned by varying the scattering particle size and density. In particular, we investigate filters suitable for use in the EXperiment for Cryogenic Large Aperture Intensity Mapping (EXCLAIM) which will field sensitive superconducting detectors in the 420--540 GHz band.\cite{2022SPIETEH,2022SPIECV,2022SPIEMR} We also target filters for application on the 2.5 THz heterodyne receiver on the Submillimeter Solar Observation Lunar Volatiles Experiment (SSOLVE).\cite{2019EPSC...13.1173L}

\section{Filter Design and Fabrication}
In general, polyimide aerogels are highly porous solids that are lightweight and mechanically robust with very low density.  They are marked by an open-celled material composition derived from a gel in which the liquid is removed while maintaining a self-assembled three-dimensional backbone.\cite{MM2012} The resulting aerogels can be formed or molded into a variety of shapes and sizes depending on desired application. The polyimide aerogel formulations we describe here exhibit mechanical flexibility and robustness.


\begin{figure}[h!]
\centering
  \includegraphics[width=0.6\textwidth]{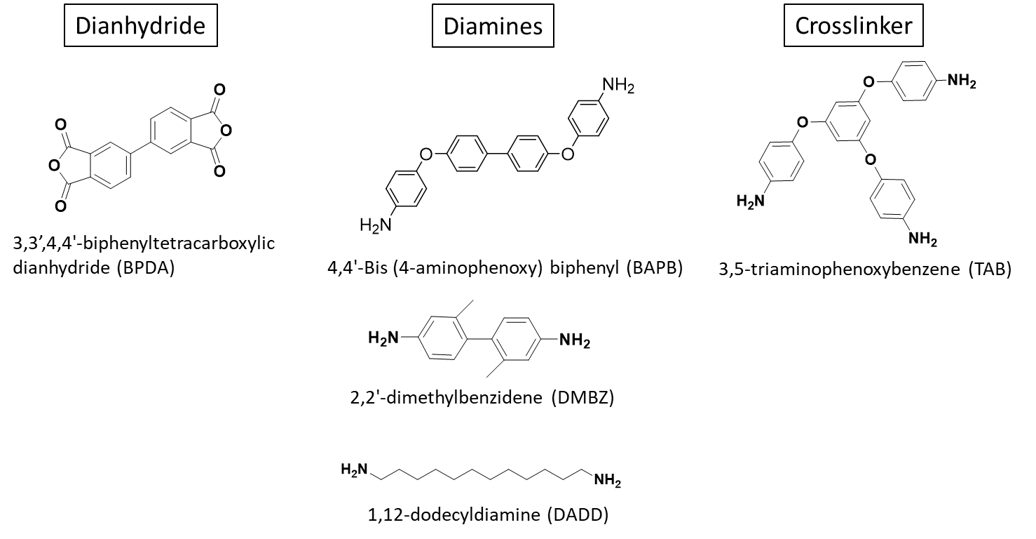}
  \caption{\label{fig:monomers} Diagrams of the monomeric precursors used in our polyimide aerogel formulations.}
\end{figure}

The aerogels investigated for this work were fabricated using biphenyl-3,3',4,4'-tetracarbooxylic dianhydride (BPDA) as dianhydride, 2,2'-dimethylbenzidine (DMBZ) as diamine in combination with 1,12-dodecyldiamine (DADD) or 4,4'-bis (4-aminophenoxy) biphenyl (BAPB), and 3,5-triaminophenoxybenzene (TAB) as cross-linker (see Fig.~\ref{fig:monomers}). Diamond scattering particles were embedded into the gel formula prior to curing and drying.\footnote{The particles were purchased from Pureon: https://pureon.com/products/diamond-powder/monocrystalline-microdiamant-mono-eco} By varying the scattering particle size and density relative to the base aerogel, we tuned the optical performance of the filters. While our group's previous work had investigated using silicon scattering particles,\cite{TEH2020} we chose to use diamond for this work due to greater commercial availability of diamond particles with known size distributions. The optical properties of diamond are also well-known, most notably low loss in the far-infrared and millimeter spectral ranges.\cite{DiamondProperties} We used a Mie scattering model to design the desired optical cut-on properties.\cite{2022SPIEKH}

\begin{figure}[h]
\centering
  \includegraphics[width=0.8\textwidth]{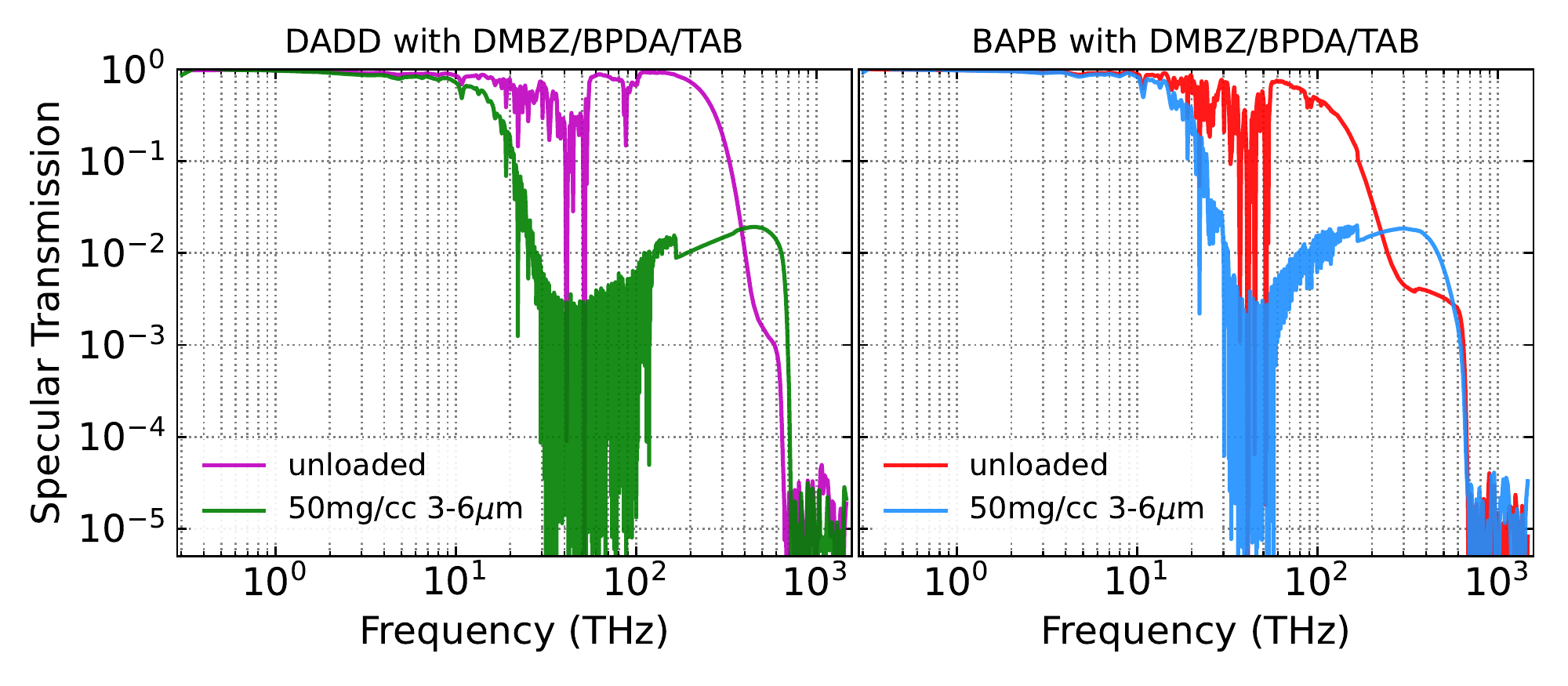}
  \caption{\label{fig:BAPBvsDADD} We compared optical performance of two base aerogel formulations with and without diamond scattering particle loading. In addition to DADD or BAPB as diamine, all four films used a combination of DMBZ (diamine), BPDA (dianhydride), and TAB (cross-linker) as the base aerogel chemistry. The measured film thicknesses were: 0.12~mm (DADD unloaded) 0.10~mm (DADD loaded), 0.11~mm (BAPB unloaded), 0.12~mm (BAPB loaded).}
\end{figure}


For ease of optical characterization, the aerogel scattering filter formulations were formed into film sheets 6--12 inches square. To form the film, the gel was film cast using a doctor blade prior to the drying process. The films shrink by approximately 10\% in all directions as they dry. Once dry, we cut 1 inch diameter circular sections of the film and use double-sided adhesive tape to mount the sections to washers for compatibility with the optical characterization facilities.

\section{Optical Characterization}

To fully characterize the optical performance of the aerogel scattering filters, we endeavor to measure the samples' specular and total hemispherical transmittance and reflectance over as broad a spectral range as possible. We use several spectrometer instruments to do this. 

The PerkinElmer LAMBDA 950 instrument is a dual-beam monochromator that operates at optical wavelengths between 200--3300 nm. We use several detectors to cover this range: a photomultiplier tube in the 200--900 nm range and a PbS photocell detector in the 800 nm--3300 nm range. An integrating sphere coated with Spectralon\footnote{https://www.labsphere.com/product/spectralon-diffuse-reflectance-material/} is used to measure total hemispherical transmittance and reflectance. The instrument's spectral accuracy is $<\pm$0.3 nm. The PerkinElmer instrument operates at ambient atmospheric pressure. 

The Bruker 125HR instrument is a versatile Fourier Transform Interferometer that operates at optical wavelengths between 500~nm and 1~mm. To span this range, we combine data taken in several instrument configurations. For wavelengths between 14~$\mu$m and 1~mm, we use a mercury arc lamp as a white light source, Mylar film beam splitter, and Si bolometer cooled to 4.2~K as a detector. For wavelengths 2--14~$\mu$m, we use a Globar white light source, KBr beam splitter, and room temperature deuterated L-alanine doped triglycine sulphate (DTGS) detector. In addition to a basic sample mount translation stage for measuring specular transmittance, we utilize  an accessory stage for measuring specular reflectance at 10$^{\circ}$ angle of incidence referenced to a gold mirror. To measure total hemispherical transmittance and reflectance for wavelengths in the range 2--20~$\mu$m, we use an accessory that couples an integrating sphere coated with a Lambertian gold surface to a DTGS detector. The spectral resolution of the Bruker 125HR is 0.0063~cm$^{-1}$, while the photometric precision is $\pm$1\%. All measurements with the Bruker are performed under vacuum (pressure below 0.4 T). 

\begin{figure}[h!]
\centering
  \includegraphics[]{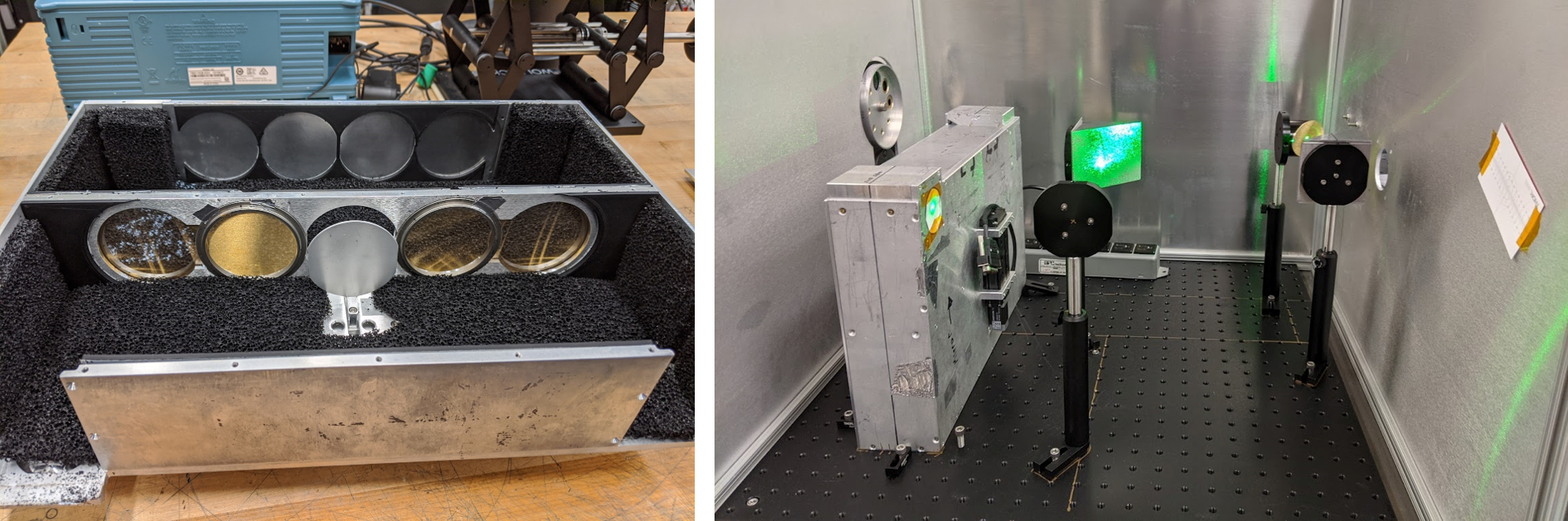}
  \caption{\label{fig:PIXIE-FTS} 
We are developing a long-wavelength FTS testbed. Left: Photo of the spectrometer unit itself (long edge measures 10~inches). Right: The FTS incorporated into our testbed with blackbody source at the left and a port for a detector at right. The green laser beam was used for initial alignment of the coupling optics.}
\end{figure}

To extend our measurement capabilities beyond 1~mm, our group has also been developing a testbed built around a compact mm-wavelength Fourier Transform Spectrometer (FTS). The FTS is designed to operate at wavelengths from 0.9--6~mm, and the design of the spectrometer itself is described in Ref. \citenum{Pan2019}. The spectrometer, shown in Fig.~\ref{fig:PIXIE-FTS}, incorporates two sets of orthogonal wire polarizing grids as beamsplitters, and machined aluminum elliptical mirrors that attach to the sides of the spectrometer enclosure. The non-optical surfaces of the enclosure are covered with foam absorber material and 3D-printed black strips surround the mirrors. We have integrated the FTS with a cavity blackbody outfitted with an optical chopper wheel to serve as a white light source.\footnote{Infrared Systems IR-564/301 https://www.infraredsystems.com/Products/blackbody564.html} We also designed a set of coupling optics to collimate the beam and allow space for an optical sample to be mounted. For the testbed, we use a liquid helium-cooled dual-bolometer system as a detector.\footnote{Infrared Laboratories: https://www.irlabs.com/products/bolometers/bolometer-systems/} The system contains two Si bolometers: one operating at 4.2~K for wavelengths below 1~mm and one operating at 1.6~K for wavelengths above 1~mm. We are working on developing the testbed into a user facility capable of measuring specular transmittance of our aerogel scattering filters.

\section{Filter Performance}

\begin{figure}[h]
\centering
  \includegraphics[width=0.7\textwidth]{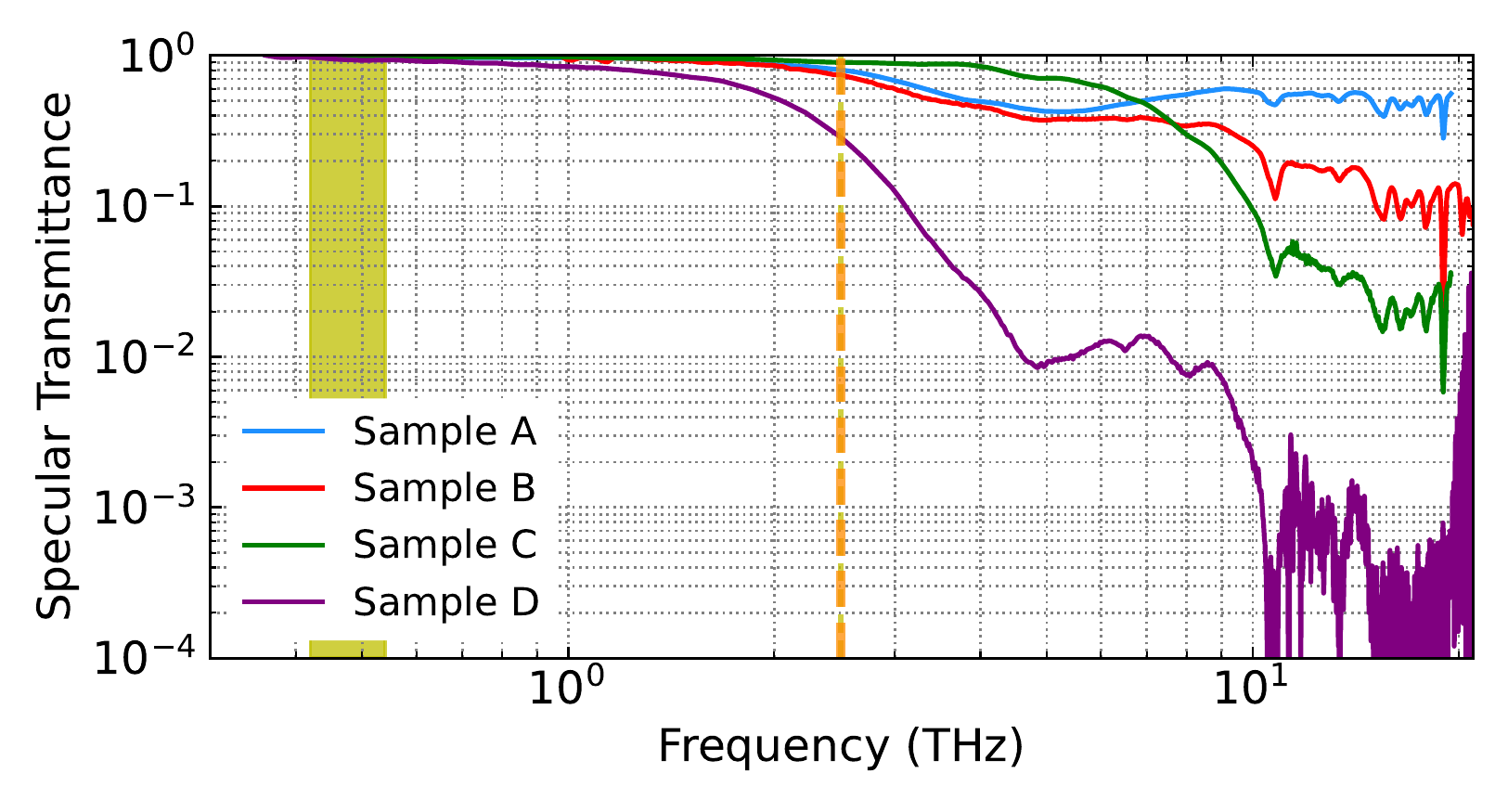}
  \caption{\label{fig:comparison-tspec} We characterize the specular transmittance of 4 formulations of aerogel filters incorporating diamond scattering particles. Sample properties are described in Table \ref{tab:ABCD}. The spectral bands of the EXCLAIM (shaded yellow) and SSOLVE (dashed orange) instruments are shown in yellow.}
\end{figure}

Here we present the optical characterization of nine samples of aerogel scattering filters, seven of which were loaded with diamond scattering particles. Fig.~\ref{fig:BAPBvsDADD} shows a direct comparison of the aerogel base formulations using either BAPB or DADD, with and without equal concentrations of diamond scattering particles. Specular transmittance measurements for four samples fabricated using the BAPB with DMBZ/BPDA/TAB base aerogel formulation are shown in Fig.~\ref{fig:comparison-tspec}. Their thicknesses and diamond particle loading concentrations are listed in Table \ref{tab:ABCD}. 

Two of these formulations were targeted towards specific mission applications. Sample D is the candidate filter formulation for use on the EXCLAIM mission. EXCLAIM requires a 50\% cut-off frequency of 2~THz and out-of-band rejection by a factor of 1000 beyond 10~THz. We measure a 50\% cut-off frequency of 2.05 THz and transmittance below $3\times10^{-3}$ between 10 and 18~THz (the increase beyond 18~THz is due to an increasing noise floor in the bolometer detector). Sample C, which is 0.08~mm thick and contains 60~mg/cc 10-20$\mu$m and 20~mg/cc 3-6$\mu$m scattering particles, is a candidate filter for use on the SSOLVE mission. SSOLVE requires transmittance above 94\% at 2.5~THz. We measure 90.3\% transmittance at 2.5~THz. For both projects' candidate formulations, the samples tested represent an initial design and fabrication cycle. With further iteration we anticipate the aerogel scattering filters will meet project requirements.

\begin{table}[h]
\caption{Aerogel sample properties} 
\label{tab:ABCD}
\begin{center}       
\begin{tabular}{|l|l|l|}
\hline
\rule[-1ex]{0pt}{3.5ex}  Sample & Thickness (mm) & Particle Loading  \\
\hline \hline
\rule[-1ex]{0pt}{3.5ex}  A & 0.09 & 50 mg/cc 40--60$\mu$m   \\
\hline
\rule[-1ex]{0pt}{3.5ex}  B & 0.16 & 40 mg/cc 40--60$\mu$m, 20 mg/cc 10--20$\mu$m, 20 mg/cc 3--6$\mu$m  \\
\hline
\rule[-1ex]{0pt}{3.5ex}  C & 0.08 & 60 mg/cc 10--20$\mu$m, 20 mg/cc 3--6$\mu$m  \\
\hline
\rule[-1ex]{0pt}{3.5ex}  D & 1.14 & 40 mg/cc 40--60$\mu$m, 20 mg/cc 10--20$\mu$m, 20 mg/cc 3--6$\mu$m  \\
\hline 
\end{tabular}
\end{center}
\end{table}

\begin{figure}[h]
\centering
  \includegraphics[width=0.7\textwidth]{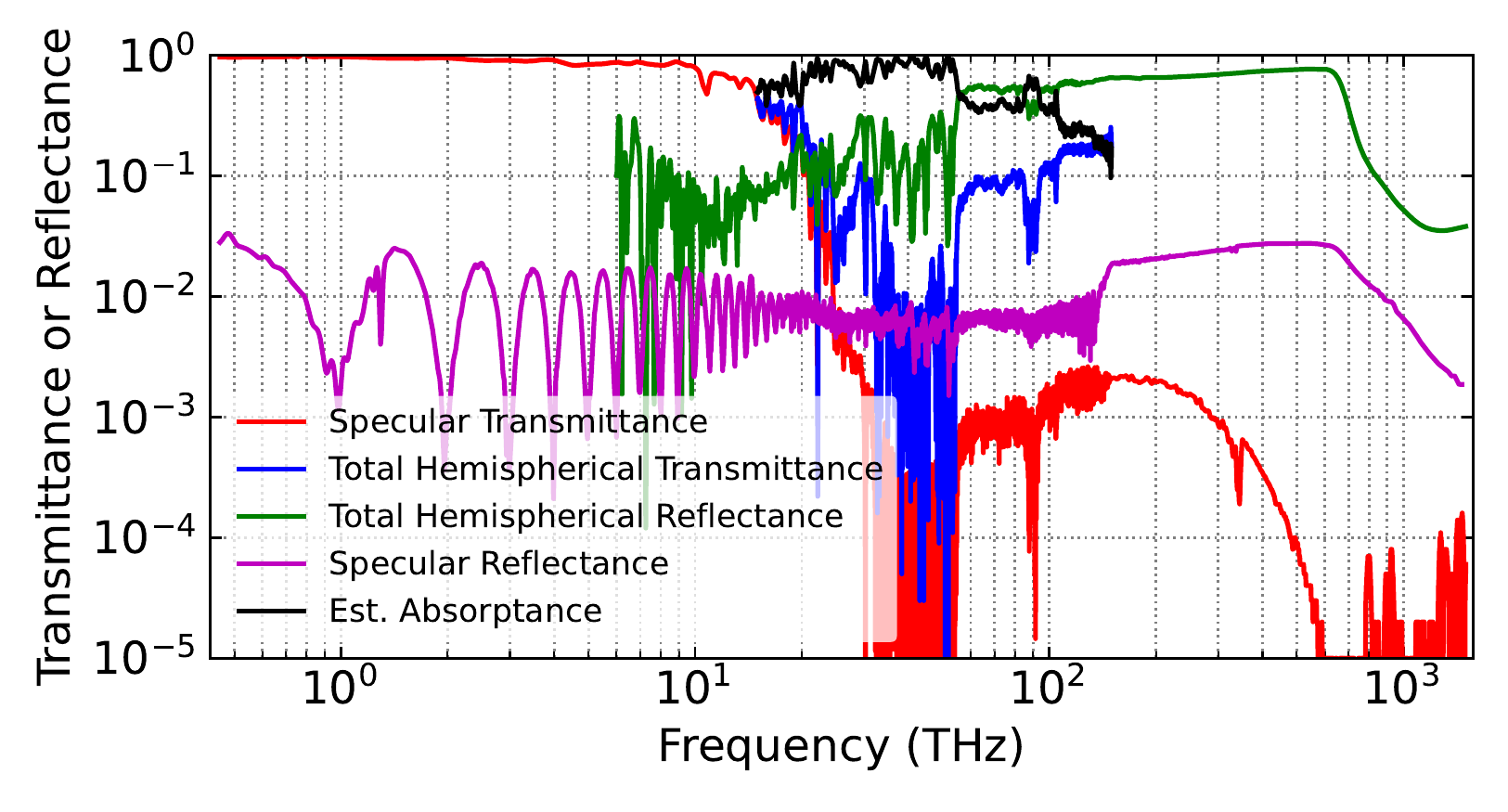}
  \caption{\label{fig:sampleD-fullcharacterization} 
We characterize the optical performance of aerogel scattering filters by measuring their specular and total hemispherical transmittance and reflectance. This sample was 0.12~mm thick and loaded with 80~mg/cc 3--6~$\mu$m diamond scattering particles.}
\end{figure}

Full optical characterization of another sample is shown in Fig.~\ref{fig:sampleD-fullcharacterization}. This sample used the BAPB with DMBZ/BPDA/TAB aerogel base formulation with 80~mg/cc loading of 3--6~$\mu$m diamond scattering particles, and was measured to be 0.12~mm thick. We estimated the absorptance of the sample by subtracting the total hemispherical reflectance and transmittance from unity. 

\subsection{Environmental Testing}

All the optical characterization data presented here were taken with the filter samples at room temperature. Our previous work has demonstrated that the polyimide aerogel material survives cryogenic cycling. Although we expect only minimal changes in optical performance as a function of temperature, we plan to characterize the filter transmittance at cryogenic temperatures since the entire EXCLAIM instrument including the optics are submerged in a liquid helium dewar. 

\begin{figure}[h]
\centering
  \includegraphics[width=0.7\textwidth]{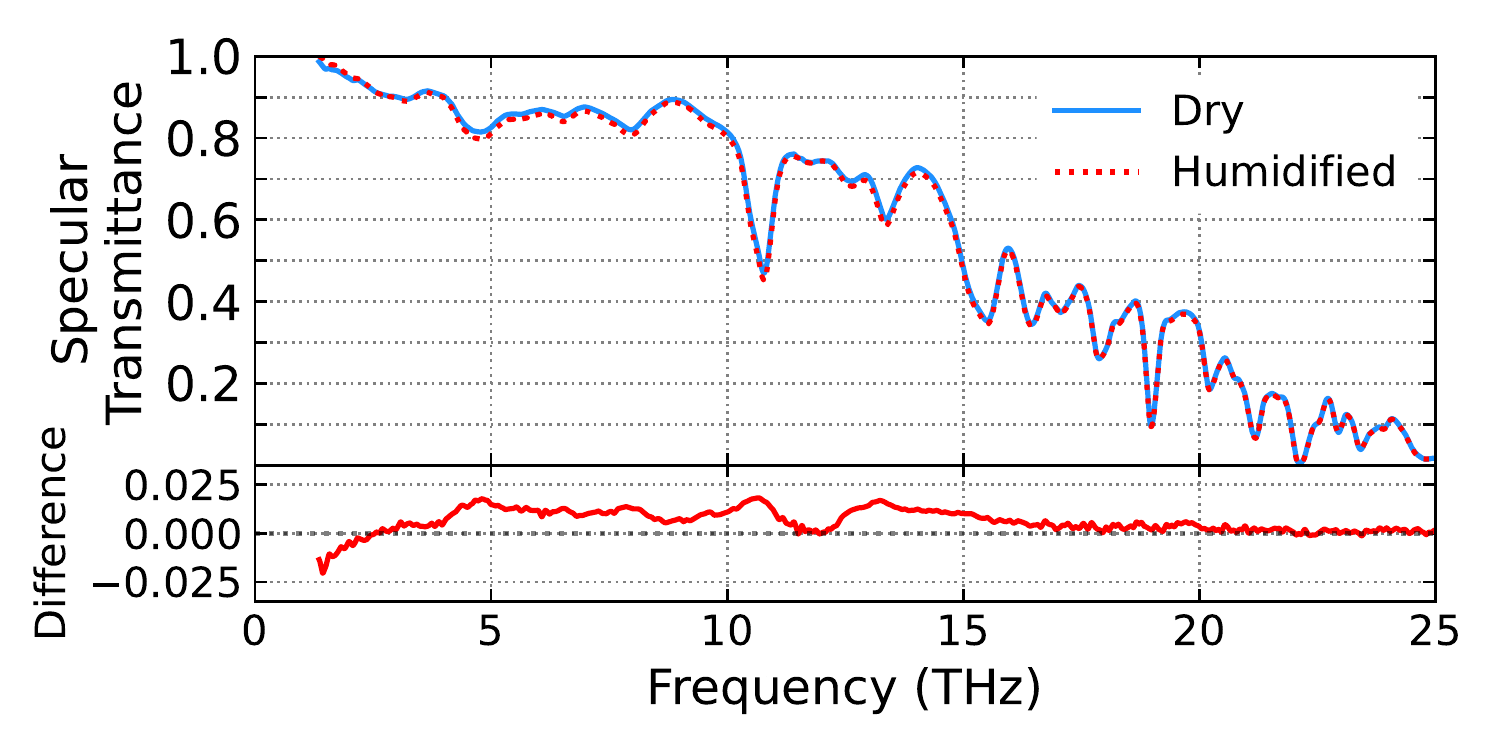}
  \caption{\label{fig:humidity} 
Specular Transmittance before and after exposure to a humid environment. This sample measured 0.12~mm thick and was fabricated using the BAPB with DMBZ/BPDA/TAB base chemistry, loaded with 50~mg/cc 3--6$\mu$m diamond scattering particles. The bottom panel shows the difference between the measurements (humidified spectrum subtracted from the dry spectrum). }
\end{figure}

We also performed an initial test to investigate the susceptibility of the aerogel filters to changes in ambient humidity. We measured the specular transmittance of a 1 inch square sample of an aerogel sample loaded with diamond scattering particles (the formulation matches the one shown in Figure \ref{fig:BAPBvsDADD} as the unloaded BAPB sample). We exposed the sample to a warm, humid environment by placing it in an oven set to 50$^\circ$C for 24 hours, and then immediately remeasured its specular transmittance. As shown in Figure \ref{fig:humidity}, the transmittance spectrum changed by less than 2\% across the 2--220~$\mu$m band we measured, and we saw no evidence of spectral features of water. 

\acknowledgments 
 
The authors gratefully acknowledge the National Aeronautics and Space Administration under grant numbers NNX14AB76A and NNH18ZDA001N-18-APRA18-0008, as well as the National Science Foundation Division of Astronomical Sciences for their support of this work under Grant Numbers 0959349, 1429236, 1636634, and 1654494. K.H acknowledges that the material is based upon work supported by NASA under award number 80GSFC17M0002. We also gratefully acknowledge support under the Goddard Internal Research and Development (IRAD) program. 

\bibliography{report} 

\begin{thebibliography}{10}

\bibitem{2006SPIE.6275E..0UA}
{Ade}, P. A.~R., {Pisano}, G., {Tucker}, C., and {Weaver}, S., ``{A review of
  metal mesh filters},'' in [{\em Society of Photo-Optical Instrumentation
  Engineers (SPIE) Conference Series}{\nolinebreak\hspace{0.1em}]},
  {Zmuidzinas}, J., {Holland}, W.~S., {Withington}, S., and {Duncan}, W.~D.,
  eds., {\em Society of Photo-Optical Instrumentation Engineers (SPIE)
  Conference Series} {\bf 6275},  62750U (June 2006).

\bibitem{2012SPIE.8452E..3IS}
{Sharp}, E.~H., {Benford}, D.~J., {Fixsen}, D.~J., {Moseley}, S.~H., {Staguhn},
  J.~G., and {Wollack}, E.~J., ``{Stray light suppression in the Goddard IRAM
  2-Millimeter Observer (GISMO)},'' in [{\em Millimeter, Submillimeter, and
  Far-Infrared Detectors and Instrumentation for Astronomy
  VI}{\nolinebreak\hspace{0.1em}]},  {Holland}, W.~S. and {Zmuidzinas}, J.,
  eds., {\em Society of Photo-Optical Instrumentation Engineers (SPIE)
  Conference Series} {\bf 8452},  84523I (Sept. 2012).

\bibitem{1986ApOpt..25..565H}
{Halpern}, M., {Gush}, H.~P., {Wishnow}, E., and {de Cosmo}, V., ``{Far
  infrared transmission of dielectrics at cryogenic and room temperatures:
  glass, Fluorogold, Eccosorb, Stycast, and various plastics},'' {\em Applied
  Optics}~{\bf 25},  565--570 (Feb. 1986).

\bibitem{2017ApOpt..56.5349M}
{Munson}, C.~D., {Choi}, S.~K., {Coughlin}, K.~P., {McMahon}, J.~J., {Miller},
  K.~H., {Page}, L.~A., and {Wollack}, E.~J., ``{Composite
  reflective/absorptive IR-blocking filters embedded in metamaterial
  antireflection-coated silicon},'' {\em Applied Optics}~{\bf 56},  5349 (June
  2017).

\bibitem{1995ApOpt..34.7254B}
{Bock}, J.~J. and {Lange}, A.~E., ``{Performance of a low-pass filter for
  far-infrared wavelengths},'' {\em Applied Optics}~{\bf 34},  7254--7257 (Nov.
  1995).

\bibitem{1974ApOpt..13..425A}
{Armstrong}, K.~R. and {Low}, F.~J., ``{Far-infrared filters utilizing small
  particle scattering and antireflection coatings.},'' {\em Applied
  Optics}~{\bf 13},  425--430 (Jan. 1974).

\bibitem{2022SPIETEH}
{Essinger-Hileman}, T.~M., {Ade}, P., {Anderson}, C., {Barlis}, A.,
  {Barrentine}, E., {Beeman}, J., {Bellis}, N., {Bolatto}, A., {Breysse}, P.,
  {Bulcha}, B., {Cataldo}, G., {Chevres Fernandez}, L.~R., {Cho}, C.,
  {Connors}, J., {Ehsan}, N., {Glenn}, J., {Golec}, J., {Hays-Wehle}, J.,
  {Hess}, L., {Jahromi}, A., {Jenkins}, T., {Kimball}, M., {Kogut}, A., {Lowe},
  L., {Mauskopf}, P., {McMahon}, J., {Mirzaei}, M., {Moseley}, H.,
  {Mugge-Durum}, J., {Noroozian}, O., {Oxholm}, T., {Parekh}, T., {Pen}, U.-L.,
  {Pullen}, A., {Rahmani}, M., {Ramirez}, M., {Roselli}, F., {Shire}, K.,
  {Siebert}, G., {Sinclair}, A., {Somerville}, R., {Stephenson}, R.,
  {Stevenson}, T., {Switzer}, E., {Timbie}, P., {Termini}, J., {Trenkamp}, J.,
  {Tucker}, C., {Visbal}, E., {Volpert}, C., {Watson}, J., {Wollack}, E.,
  {Yang}, S., and {Yung}, A., ``{EXCLAIM: The EXperiment for Cryogenic
  Large-Aperture Intensity Mapping},'' in [{\em Millimeter, Submillimeter, and
  Far-Infrared Detectors and Instrumentation for Astronomy
  XI}{\nolinebreak\hspace{0.1em}]},  {Gao}, J.-R. and {Zmuidzinas}, J., eds.,
  {\em Society of Photo-Optical Instrumentation Engineers (SPIE) Conference
  Series} {\bf 12190} (2022).

\bibitem{2022SPIECV}
{Volpert}, C.~G., {Barrentine}, E.~M., {Mirzaei}, M., {Barlis}, A., {Bolatto},
  A.~D., {Bulcha}, B.~T., {Cataldo}, G., {Connors}, J.~A., {Costen}, N.,
  {Ehsan}, N., {Essinger-Hileman}, T., {Glenn}, J., {Hays-Wehle}, J.~P.,
  {Hess}, L.~A., {Kogut}, A.~J., {Moseley}, H., {Mugge-Durum}, J., {Noroozian},
  O., {Oxholm}, T.~M., {Rahmani}, M., {Stevenson}, T., {Switzer}, E.~R.,
  {Watson}, J., and {Wollack}, E.~J., ``{Developing a new generation of
  integrated micro-spec far-infrared spectrometers for the experiment for
  cryogenic large-aperture intensity mapping (EXCLAIM)},'' in [{\em Space
  Telescopes and Instrumentation 2022: Optical, Infrared, and Millimeter
  Wave}{\nolinebreak\hspace{0.1em}]},  {Coyle}, L.~E., {Matsuura}, S., and
  {Perrin}, M.~D., eds., {\em Society of Photo-Optical Instrumentation
  Engineers (SPIE) Conference Series} {\bf 12180} (2022).

\bibitem{2022SPIEMR}
{Rahmani}, M., {Barlis}, A., {Barrentine}, E.~M., {Brown}, A.~D., {Bulcha},
  B.~T., {Cataldo}, G., {Connors}, J., {Ehsan}, N., {Essinger-Hileman}, T.~M.,
  {Grant}, H., {Hays-Wehle}, J., {Hsieh}, W.-T., {Mikula}, V., {Moseley},
  S.~H., {Noroozian}, O., {Oxholm}, T.~R., {Quijada}, M.~A., {Patel}, J.,
  {Stevenson}, T.~R., {Switzer}, E.~R., {Tucker}, C., {U-Yen}, K., {Volpert,
  Carolyn}, and {Wollack}, E.~J., ``{Optical characterization and testbed
  development for $\mu$-Spec integrated spectrometers},'' in [{\em Space
  Telescopes and Instrumentation 2022: Optical, Infrared, and Millimeter
  Wave}{\nolinebreak\hspace{0.1em}]},  {Coyle}, L.~E., {Matsuura}, S., and
  {Perrin}, M.~D., eds., {\em Society of Photo-Optical Instrumentation
  Engineers (SPIE) Conference Series} {\bf 12180} (2022).

\bibitem{2019EPSC...13.1173L}
{Livengood}, T.~A., {Anderson}, C.~M., {Bradley}, D.~C., {Bulcha}, B.~T.,
  {Chin}, G., {Hewagama}, T., {Jamison-Hooks}, T.~L., and {Racette}, P.~E.,
  ``{Submillimeter Solar Observation Lunar Volatiles Experiment (SSOLVE)},'' in
  [{\em EPSC-DPS Joint Meeting 2019}{\nolinebreak\hspace{0.1em}]},   {\bf
  2019},  EPSC--DPS2019--1173 (Sept. 2019).

\bibitem{MM2012}
Meador, M. A.~B., Malow, E.~J., Silva, R., Wright, S., Quade, D., Vivod, S.~L.,
  Guo, H., Guo, J., and Cakmak, M., ``Mechanically strong, flexible polyimide
  aerogels cross-linked with aromatic triamine,'' {\em ACS applied materials \&
  interfaces}~{\bf 4}(2),  536--544 (2012).
\newblock Publisher: ACS Publications.

\bibitem{TEH2020}
Essinger-Hileman, T., Bennett, C.~L., Corbett, L., Guo, H., Helson, K.,
  Marriage, T., Meador, M. A.~B., Rostem, K., and Wollack, E.~J., ``Aerogel
  scattering filters for cosmic microwave background observations,'' {\em
  Applied Optics}~{\bf 59},  5439--5446 (June 2020).
\newblock Publisher: Optica Publishing Group.

\bibitem{DiamondProperties}
Thomas, M.~E. and Tropf, W.~J., ``Optical properties of diamond,'' in [{\em
  Window and {Dome} {Technologies} and {Materials}
  {IV}}{\nolinebreak\hspace{0.1em}]},   {\bf 2286},  144--151, SPIE (Sept.
  1994).

\bibitem{2022SPIEKH}
{Helson}, K.~R., {Arseneau}, S., {Barlis}, A., {Bennett}, C.~L.,
  {Essinger-Hileman}, T., {Guo}, H., {Marriage}, T., {Tokarz}, A.~E., {Vivod},
  S.~L., and {Wollack}, E.~J., ``{Novel infrared-blocking aerogel scattering
  filters and their applications in astrophysical and planetary science
  observations},'' in [{\em Millimeter, Submillimeter, and Far-Infrared
  Detectors and Instrumentation for Astronomy XI}{\nolinebreak\hspace{0.1em}]},
   {Gao}, J.-R. and {Zmuidzinas}, J., eds., {\em Society of Photo-Optical
  Instrumentation Engineers (SPIE) Conference Series} {\bf 12190} (2022).

\bibitem{Pan2019}
Pan, Z., Pan, Z., Pan, Z., Liu, M., Liu, M., Thakur, R.~B., Benson, B.~A.,
  Benson, B.~A., Benson, B.~A., Fixsen, D.~J., Fixsen, D.~J., Goksu, H., Rath,
  E., Meyer, S.~S., Meyer, S.~S., Meyer, S.~S., and Meyer, S.~S., ``Compact
  millimeter-wavelength {Fourier}-transform spectrometer,'' {\em Applied
  Optics}~{\bf 58},  6257--6267 (Aug. 2019).
\newblock Publisher: Optica Publishing Group.

\end{thebibliography}
\bibliographystyle{spiebib} 

\end{document}